# HAIR AND SCALP DISEASE DETECTION USING DEEP LEARNING

Kavita Sultanpure[1,] Bhairavi Shirsath[2], Bhakti Bhande[3], Harshada Sawai[4], Srushti Gawade[5], Suraj Samgir[6]

[1-6]*Dept. of Information Technology*
*Vishwakarma Institute of Technology Pune, India*

kavita.sultanpure1@vit.edu[1], bhairavi.shirsat23@vit.edu[2], bhakti.bhande23@vit.edu[3],
harshada.sawai23@vit.edu[4], srushti.gawade23@vit.edu[5], suraj.samgir23@vit.edu[6]

**Abstract:** In recent years, there has been a notable advancement in the integration of healthcare and technology, particularly evident in the field of medical image analysis. This paper introduces a pioneering approach in dermatology, presenting a robust method for the detection of hair and scalp diseases using state-of-the-art deep learning techniques. Our methodology relies on Convolutional Neural Networks (CNNs), well-known for their efficacy in image recognition, to meticulously analyze images for various dermatological conditions affecting the hair and scalp. Our proposed system represents a significant advancement in dermatological diagnostics, offering a non-invasive and highly efficient means of early detection and diagnosis. By leveraging the capabilities of CNNs, our model holds the potential to revolutionize dermatology, providing accessible and timely healthcare solutions. Furthermore, the seamless integration of our trained model into a web-based platform developed with the Django framework ensures broad accessibility and usability, democratizing advanced medical diagnostics. The integration of machine learning algorithms into web applications marks a pivotal moment in healthcare delivery, promising empowerment for both healthcare providers and patients. Through the synergy between technology and healthcare, our paper outlines the meticulous methodology, technical intricacies, and promising future prospects of our system. With a steadfast commitment to advancing healthcare frontiers, our goal is to significantly contribute to leveraging technology for improved healthcare outcomes globally. This endeavor underscores the profound impact of technological innovation in shaping the future of healthcare delivery and patient care, highlighting the transformative potential of our approach.

## I. INTRODUCTION:

A scalp self-diagnosis algorithm utilizing smart device microscopy is developed to mitigate alopecia areata's increasing cases. Through image processing techniques, it extracts hair loss features (HLF), including follicle count, hair width, and integration of various hair types, aiding in assessing hair loss progression and potentially contributing to early intervention strategies.[1]

This study introduces a cost-effective method for scalp condition monitoring using pre-trained image processing techniques, aiming to address the limitations of traditional expensive and uncomfortable treatments. By extracting attributes like color, texture, and shape from scalp images and employing Support Vector Machine (SVM) classification, it achieves an 85% accuracy rate in categorizing scalp conditions (dandruff, normal, alopecia areata). This automated approach holds promise for facilitating timely treatment decisions based on accurate scalp condition assessment.[2]

This study employs deep learning techniques, specifically a 2D convolutional neural network (CNN), to predict scalp conditions like alopecia, psoriasis, and folliculitis with high accuracy (91.1% validation, 96.2% training). Despite challenges such as limited research and diverse image sources, preprocessing methods and a dataset of 150 photos were utilized to reduce error rates, providing valuable insights for both doctors and patients in early symptom detection and diagnosis.[3]

This study evaluates the use of artificial neural networks, specifically a feedforward neural network with backpropagation, for identifying human alopecia with a reliable accuracy of 91%. By providing clinical specialists with a second opinion, the proposed model aids in making informed diagnostic decisions, addressing the challenges of expanding populations, environmental factors, and the need for accurate disease identification in dynamic symptomatology. [4]

Scalp dermoscopy, or trichoscopy, offers a noninvasive method for evaluating hair loss by magnifying scalp skin and hair. Utilizing manual or video dermoscopes with varying magnifications, up to ×1,000, trichoscopy aids in diagnosing various conditions like alopecia areata and telogen effluvium, minimizing the need for scalp biopsies and facilitating therapy monitoring for patients' favorable responses.[5]

## II. LITERATURE SURVEY:

Researchers employed scalp images to identify alopecia areata features. They introduced a novel trichoscopy method utilizing computer vision techniques like grid line selection and eigenvalue analysis. This innovative system aimed to diagnose alopecia areata by automating hair loss feature identification, marking a significant fusion of machine vision and image processing for detection [1].

In dermatology, machine learning methods like SVM and KNN have been successful in diagnosing and predicting conditions through scalp analysis systems. These techniques are utilized for classifying scalp images, including conditions like dandruff, employing SVM, KNN, and decision trees. However, despite their effectiveness, none of these methods have been applied specifically to human hair images yet 14 to 20. [2].A framework for classifying hair was suggested by Kapoor I and Mishra, which distinguishes healthy hair from alopecia areata using features from hair photos, such as color, texture, and shape. They employed SVM with KNN, achieving accuracies of 91.4% and 88.9%, respectively, indicating successful classification. Future advancements could involve integrating deep learning techniques like CNN into this framework for enhanced performance.[3]

According to Lacarrubba et al. [4], scalp hair dermoscopy, sometimes referred to as scalp image microscopy, is a helpful technique for treating problems and keeping an eye on symptoms related to the skin and scalp hair. Trichoscopy, which includes dermoscopy of the scalp and hair, was utilised by Rudnicka et al. to detect illnesses of the scalp and hair, including trichotillomania, tinea capitis, and alopecia areata. Kim et al. assessed the state of the hair and scalp using measurements and analysis of microscope images.

Lee and Yang [5] developed an intelligent hair and scalp analysis system that extracted picture features and parameters using a web camera, a microscope image sensor, and the Norwood–Hamilton scaling model in order to assess the state of customers' scalp hair. This experiment was conducted using the Nvidia Jetson TK1 development platform, which has the ability to conduct basic scalp hair self-diagnosis.

III. METHODOLOGY

*Method of Data Analysis:*

1. Image Processing :
Image processing involves manipulating images to enhance quality, often starting with pre-processing methods like noise reduction, equalization, reflection removal, and intensity adjustments to improve visual clarity. These techniques, utilizing low pass filters to address high-frequency noise, prepare images by mitigating distortions and enhancing specific visual elements, facilitating human perception and providing better input for automated image processing systems.

2. Neural Network training:

Deep learning, an evolution of machine learning, employs convolutional neural networks (CNNs) to simulate human learning, enhancing tasks like computer vision and speech recognition. CNNs utilize hierarchical feature extraction and parameter sharing, exhibiting lower complexity compared to fully connected networks. Nonlinearities introduced by activation functions like ReLU mitigate saturation issues, enhancing accuracy at low computational cost. Pooling layers reduce parameters and dimensionality, governed by hyperparameters like filter size and stride. Fully connected layers enable advanced reasoning by connecting neurons and facilitating estimation through affine transformations.

IV. PROPOSED SYSTEM:

A novel medicinal plant identification approach uses photos of leaves from various perspectives, combining distinctive morphological characteristics to increase identification rates. Utilizing Dense Net CNNs, the system can determine if a leaf belongs to a medicinal plant, providing details on its properties, names, and medicinal uses with improved efficiency and reduced valuation loss through feature reuse and propagation.

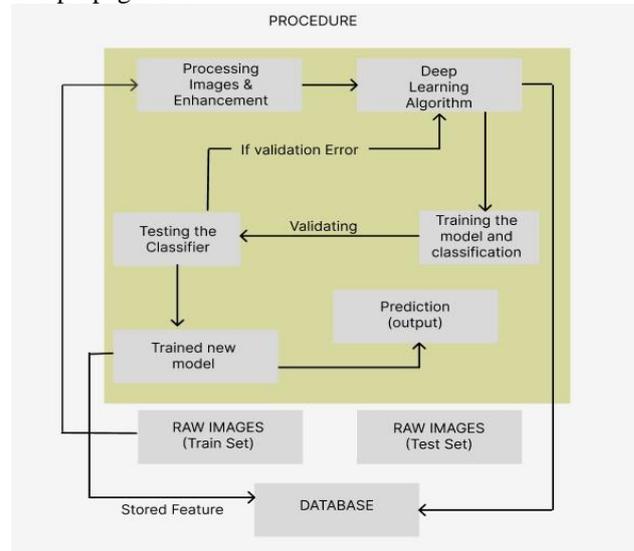

Fig 1. System Architecture

V. ALGORITHM:

1. Import necessary Libraries-
   Utilize TensorFlow for the reference.
Import the requisite TensorFlow modules, incorporating layers, models, and pyplot.

2. Set Constants
   Decide on 32 as the batch size.
   Decide on 256 as the image size.
   In the case of RGB photos, set the number of channels to 3
   Assign 50 epochs to the training set.

3. Load Dataset
   Indicate the dataset's route.
   To load dataset, use the image dataset_from_directory().

4. Visualize Dataset
   Present a portion of the loaded dataset for your perusal.

5. Split Dataset
   Utilize the dataset to create training, validation, and test sets. It is recommended to allocate 10% for testing, 10% for validation, and 80% for training.

6. Preparation of Dataset
   Cache, shuffle, and prefetch the test, validation, and training datasets to improve training speed.

7. Construct the Model
   Convolutional Neural Network (CNN) models should constructed.
   Give the layers for scaling and normalization.

8. Augmenting Data
   Employ data augmentation techniques to raise the model's efficiency and generalizability.

9. Utilize augmentation of data
   Apply the data augmentation procedures to the training dataset.

10. Put together the model. Utilize the Adam optimizer accuracy as a metric, and the Sparse Categorical Cross entropy loss function to build the model.

11. Model Train
At the conclusion of each epoch, validate the model using the validation set after training it on the training dataset.

12. Assess the Model
Utilizing the test dataset, evaluate the trained model's performance.

13. Store the Model
Save the trained model to a designated directory so that it can be used or deployed later.

## VI. COMPARATIVE ANALYSIS:

Table 1: Comparative table for Accuracy

| System | Accuracy (%) |
|---|---|
| Proposed System | 99% |
| S. Seo and J. Park [1] | 91.4% |
| S. Ibrahim, Z. A. Noor Azmy, N. N. Abu Mangshor [2] | 96.2% |

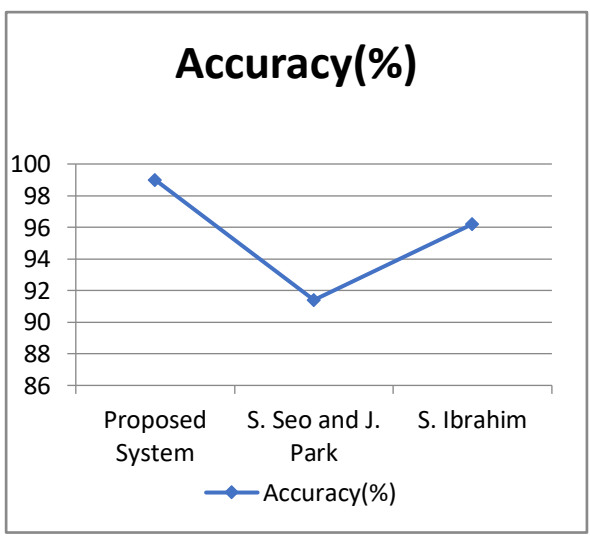

Fig2. shows the system and S. Ibrahim, Z. A. Noor Azmy, N. N. Abu Mangshor [2] demonstrates the highest accuracy (99%) among the systems.

Table 2: Comparative table for Precision

| System | Precision |
|---|---|
| Proposed System | 99% |
| S. Seo and J. Park [1] | 81% |
| S. Ibrahim, Z. A. Noor Azmy, N. N. Abu Mangshor [2] | 80% |

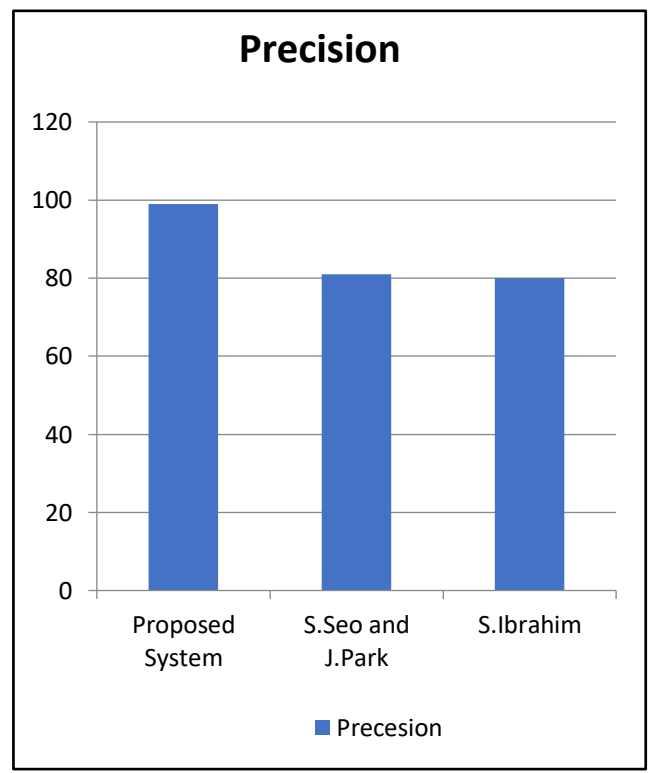

Fig3. shows the proposed system and S. Seo and J. Park [1] demonstrates the highest precision (0.99) among the system.

Table 3. Comparative Table for Fscore

| System | Fscore |
|---|---|
| Proposed System | 99% |
| S. Seo and J. Park [1] | 86% |
| S. Ibrahim, Z. A. Noor Azmy, N. N. Abu Mangshor [2] | 100% |

Table 4. Comparative table for Recall

| System | Recall |
|---|---|
| Proposed System | 0.99 |
| S. Seo and J. Park [1] | 0.91 |
| S. Ibrahim, Z. A. Noor Azmy, N. N. Abu Mangshor [2] | 0.8 |

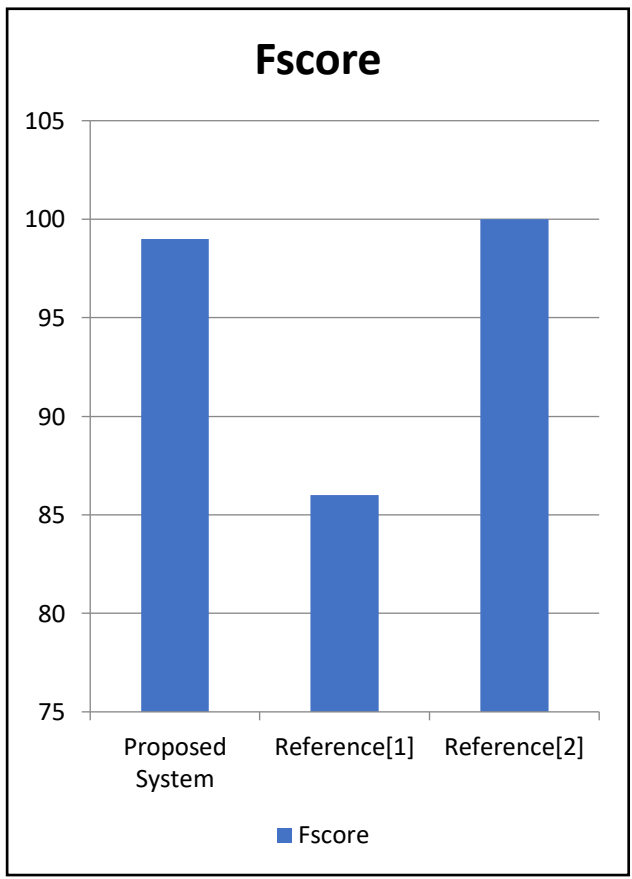

Fig 4. shows the table and graph above compare the Precision of different hair and scalp disease identification systems using decision trees.

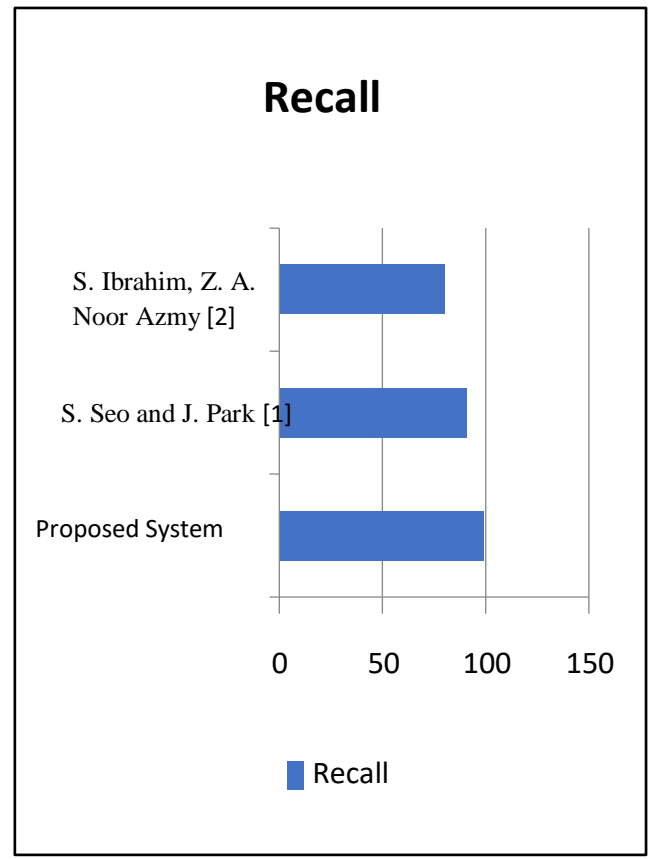

Fig 5. Shows the table and graph above compare the Recall of different hair and scalp disease detection system

## VII. TABLES AND GRAPHS:

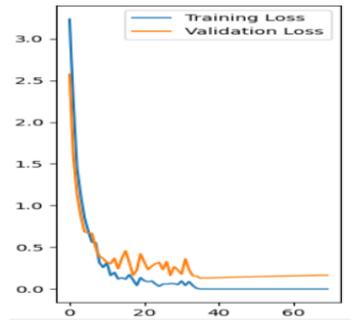

Fig 6. Training and Validation Loss

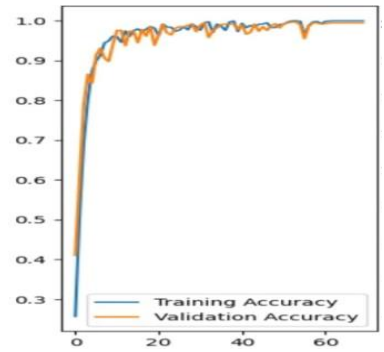

Fig 7. Training and Validation Accuracy

## VIII. RESULT:

| Sr.No. | Input | Output | Remark |
|---|---|---|---|
| 1) | 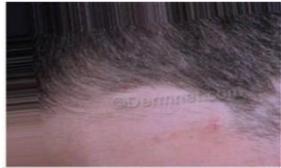 | 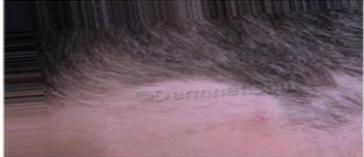 | Actual: Alopecia Areata<br>Predicted: Alopecia Areata<br>Confidence: 100.0% |
| 2) | 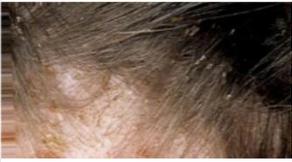 | 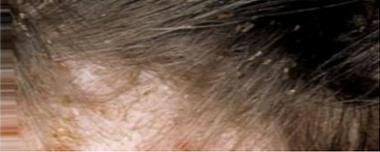 | Actual: Head Lice<br>Predicted: Head Lice<br>Confidence: 100.0% |
| 3) | 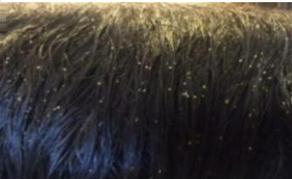 | 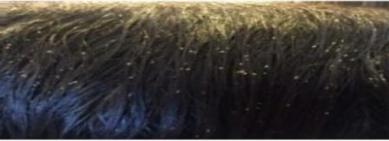 | Actual: Head Lice<br>Predicted: Head Lice<br>Confidence: 100.0% |
| 4) | 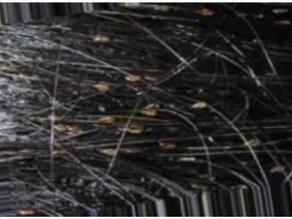 | 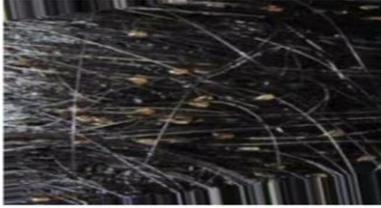 | Actual: Head Lice<br>Predicted: Head Lice<br>Confidence: 100.0% |

## IX. SCOPE OF RESEARCH:

Accuracy Enhancement: Ongoing research will concentrate on improving the machine learning model's ability to accurately identify hair and scalp disease.

Database Expansion: A wider range of hair and scalp will be included as the dataset is progressively increased.

Mobile Application: The system's capabilities can be expanded to create a portable hair and scalp identification app.

Community Participation: Users can encourage participation in the identification of hair and scalp disease by adding photographs to the database.

## X. FUTURE SCOPE:

Improved Model Performance Discuss potential avenues for enhancing the performance and accuracy of deep learning models in disease detection. Expansion to Other Conditions Explore the possibilities of leveraging similar methodologies to detect and diagnose other dermatological conditions. Exploration of AI Driven Diagnostics Delve into the potential impact of artificial intelligence on the overall field of medical diagnostics.

## XI. CONCLUSION

Advancements in Healthcare Discuss how deep learning has the potential to revolutionize diagnostics and improve patient care. Collaborations & Partnerships Encourage collaborations and partnerships between researchers, medical professionals, and technology experts to drive innovation in hair and scalp disease detection. The Future of Diagnostics Look ahead to the future, where deep learning and AI will continue to play a significant role in transforming healthcare practices.